\documentclass[twocolumn,prb,aps,showpacs,ams]{revtex4}
\usepackage{latexsym,graphicx,epsfig}
\def \ba {\begin{eqnarray}}
\def \ea {\end{eqnarray}}

\def \ua {\uparrow}
\def \da {\downarrow}
\def \sech {\rm sech}

\begin{document}

\title{Interplay between spin density wave and $\pi$ phase shifted superconductivity
in the Fe pnictide superconductors}

\author{Nayoung Lee}
\affiliation{Department of Physics and Institute for Basic
Science Research, SungKyunKwan University, Suwon 440-746, Korea.
\\}

\author{Han-Yong Choi\email[To whom the
correspondences should be addressed:~]{ hychoi@skku.edu}}
\affiliation{Department of Physics and Institute for Basic
Science Research, SungKyunKwan University, Suwon 440-746, Korea.
\\}


\begin{abstract}

We explore if the phase separation or coexistence of the spin
density wave (SDW) and superconductivity (SC) states has any
relation to the incommensurability of the SDW in the Fe pnictide
superconductors. A systematic method of determining the phase
separation or coexistence was employed by computing the anisotropy
coefficient $\beta$ from the the 4th order terms of the
Ginzburg--Landau (GL) expansion of the free energy close to the
tricritical/tetracritical point. It was complemented by the
self-consistent numerical iterations of the gap equations to map
out the boundaries between the phase separation and coexistence of
the SDW and SC phases, and between commensurate (C) and
incommensurate (IC) SDW in the temperature--doping plane. Our
principal results for the sign reversed $s$-wave pairing SC, in
terms of the multicritical temperature, $T_c$, the phase
separation/coexistence boundary between the SDW and SC, $T^*$, and
the boundary between C/IC SDW, $T_M^*$, are: (a) IC-SDW and SC
coexist for $T_c < T^*$ and phase separate otherwise, (b) SDW
takes the C form for $T_c>T_M^*$ and IC form for $T_c<T_M^*$, and
(c) the thermodynamic first order phase transition intervenes in
between the C-SDW and IC-SDW boundary for large $T_M^0$, where
$T_M^0$ is the SDW transition temperature at zero doping,
$T^*=0.35 ~T_M^0$ and $T_M^*=0.56\ T_M^0$. The intervention makes
the phase diagram more complicated than previously reported. By
contrast no coexistence was found for the equal sign pairing SC.
These results will be compared with the experimental reports in
the Fe pnictide superconductors.

\end{abstract}



\maketitle

\section{Introduction}

The alluring prospect of opening a key window to understanding
the mechanism of high temperature superconductivity has attracted
enormous research activities in the iron based
pnictides.\cite{Kamihara08jacs,ChenGF08prl,ChenXH08nature} A
widespread school of thought regarding the pairing interaction
maintains that magnetic fluctuations are intimately involved for
the superconductivity. This view seems natural from the overall
phase diagram of the pnictides in the temperature and doping
plane. The superconductivity (SC) emerges out of the parent
antiferromagnetic (AF) state as the static AF order is suppressed
and charge carriers are introduced through the electron or hole
doping in common with the cuprate high temperature
superconductors.\cite{Fang0903.2418} This view is further
supported by the neutron scattering investigations showing a
resonance at the wave vector related to the AF order coincident
with the onset of superconductivity.\cite{Christianson08nature}
%
%

The intimacy of SC and AF or spin density wave (SDW) states is a
common feature among the unconventional superconductors like the
cuprates, pnictides, and heavy fermion
superconductors.\cite{Uemura09naturemat} It will, therefore, be
important to understand the interplay between the SC and SDW
orders. For the cuprates, it is perhaps one of the best
established observations that the SC emerges out of AF parent
state as the doping is introduced.\cite{Sanna04prl} For the heavy
fermion Ce115 compounds, the interplay between them has been
studied actively for CeRhIn$_5$ and CeCoIn$_5$. For CeRhIn$_5$,
for instance, the AF and SC phases coexist under pressure at zero
magnetic field.\cite{Knebel09pssb,Park06nature} Within the
coexisting dome, the AF changes from incommensurate (IC) to
commensurate (C) SDW as the pressure is
increased.\cite{Yashima09prb}

Experimental investigations on the interplay between the SC and
SDW for the pnictides report diverse results with regard to the
pnictide families and the dopants. For the 1111 family, Luetkens
$et~al.$\cite{Luetkens09naturemat} reported from the muon spin
relaxation ($\mu$SR) and M\"{o}ssbauer spectroscopy on the
LaFeAsF$_x$O$_{1-x}$ (La1111) compounds that the magnetic state
disappears and superconductivity emerges abruptly as the doping
$x$ is increased. For Sm1111 system, Drew $et~al.$ found that the
AF and SC regions coexist which could be due to phase
separation.\cite{Drew09naturemat} The neutron scattering
measurements for the Ce1111 compounds reported a magnetic phase
diagram similar to La1111.\cite{Zhao08naturemat} For the 122
family, Laplace $et~al.$\cite{Laplace09prb} reported that the
IC-SDW coexists with the SC on the atomic scale in
Ba(Fe$_{1-x}$Co$_x$)$_2$As$_2$ compounds by measuring $^{75}$As
nuclear magnetic resonance (NMR) and susceptibility. Julien
$et~al.$\cite{Julien09epl} contrasted K and Co doped 122 compounds
by also measuring $^{75}$As NMR and found that the SC and SDW
phase separate for Ba$_{0.6}$K$_{0.4}$Fe$_2$As$_2$ but
microscopically coexist for Ba(Fe$_{1-x}$Co$_x$)$_2$As$_2$ in
accord with Laplace $et~al.$ Parker $et~al.$\cite{Parker10prl}
reported from the neutron and muon experiments on
NaFe$_{1-x}$M$_x$As (M=Co, Ni) that the SC and SDW coexist on
atomic level.



Theoretically the interplay between SC and SDW for the Fe pnictide
superconductors was studied in a simple two band model by many
groups.\cite{Vorontsov09prb,Vorontsov10prb,Cvetkovic09epl,Parker09prb,Fernandes10prb}
They reported that the coexistence of SC and SDW is possible when
SDW is IC for the $\pi$ phase shifted pairing. In this paper we
also take the two band model to study the interplay between the SC
and SDW. We make the Ginzburg-Landau (GL) expansion of the free
energy close to the multicritical point for IC as well as C-SDW
states. The nature of transitions was then determined from the 4th
order terms, that is, whether they are continuous or discontinuous
and whether the two orders coexist or phase separate. The phase
separation or coexistence was determined by computing the
anisotropy coefficient $\beta$ from the 4th order terms of the GL
expansion for general IC SDW states. The superconducting state was
modeled in terms of the sign reversed two band SC
theory.\cite{Mazin08prl,Kuroki08prl,Bang08prb} The $0<\beta<1$ or
not means phase coexistence or separation. See Eqs.\ (\ref{beta})
and (\ref{coexist}) below. The multicritical temperature $T_c$
which equals to the superconducting critical temperature in the
simple model we took is set by the interaction $\lambda_S$. The
SDW transition temperature at zero doping $T_M^0\equiv
T_M(\delta=0)$ is set by $\lambda_M$. $\delta$ is the deviation
from the perfect nesting and may be tuned by doping or pressure.

When $\lambda_M$ is only slightly larger than $\lambda_S$, then
$T_M^0/T_c \gtrsim 1$, and the multicritical point occurs at a
small doping. The deviation from the perfect nesting is then small
and the SDW takes the commensurate form. The computation of
$\beta$ yields $\beta \geq 1$ which means that the C-SDW and SC
phase separate. As the ratio of $T_M^0/T_c$ increases,
$T_M(\delta)$ equals to $T_c$ at a larger $\delta$, and the SDW
becomes IC for $T_c< T_M^*$. We use the notations that $T_M^*$
represents the boundary between C-SDW and IC-SDW determined by
Eq.\ (\ref{eqtm}) and $T^*$ represents the phase
separation/coexistence boundary between SDW and SC determined by
Eq.\ (\ref{coexist}) below. $T_M^*>T^*$ in our model. The IC-SDW
and SC phases remain separated for $T_c>T^*$. For $T_c<T^*$, the
$\beta$ becomes $0<\beta<1$ and the IC-SDW and SC phases coexist.
We obtain $T_M^*=0.56\ T_M^0$ in agreement with Ref.\
\cite{Vorontsov09prb}, and $T^*=0.35\ T_M^0$. See Eqs.\
(\ref{tmstar}) and (\ref{tstar}) below.

The SDW transition as $T$ is reduced is continuous for almost all
of the parameter space. But, as the doping at the multicritical
point is increased, the 4th order term of the SDW order $\beta_M$
of Eqs.\ (\ref{free4}) and (\ref{betaM}) becomes negative and the
first order transition intervenes in between the C-SDW and IC-SDW
boundary. The intervention of the discontinuous transition makes
the phase boundary more complicated. See the Fig.\ 1 and
discussions below for details. Note that this 1st order transition
is as the temperature is lowered as presented in Fig.\ 3. On the
other hand, the 1st order transition between SDW and SC as
$\delta$ is varied below $T_c$ corresponds to the phase
separation as shown in Fig.\ 2.

The GL expansion was combined with the self-consistent numerical
iterations of the gap equations to map out the boundary between
the phase separation and coexistence of SDW and SC in the plane
of $T$ and $\delta$ for the pnictide superconductors. See the the
figures 2--4.

\begin{figure}
\includegraphics[scale=0.8]{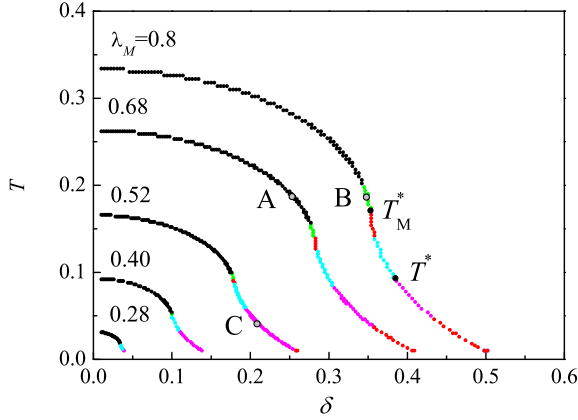}
\caption{The plot of $T_M(\delta)$ for $\lambda_M =$ 0.28, 0.4,
0.52, 0.68, and 0.80, and $T_M^*$ and $T^*$. The unit of energy
is $\omega_c$. ($\omega_c=1$). $T_M^*$ is the boundary between
$q=0$ (C) and $q\ne 0$ (IC) SDW from solving Eq.\ (\ref{eqtm}).
$T^*$ is the boundary between the phase separation and
coexistence between SDW and SC determined by Eq.\
(\ref{coexist}). For a given $\lambda_M$, in terms of the
multicritical temperature $T_c$, the region of $T_M^0 >T_c>T_M^*$
(black) corresponds to the phase separation between C-SDW and SC,
$T_M^*
>T_c>T^*$ (cyan) to the phase separation between IC-SDW and SC,
and $T_c<T^*$ (pink) corresponds to the coexistence between
IC-SDW and SC. As $\lambda_M$ is increased the region of the
discontinuous thermodynamic transitions intervenes around $T_M^*$
and in the low temperature region as indicated by the green and
red lines. The points marked by A, B, and C correspond to the
representative cases of the phase separated C-SDW and SC, phase
separated discontinuous C-SDW and SC, and coexisting IC-SDW and
SC phases, respetively. The A, B, C cases are presented in more
detail in Figs.\ 2, 3, and 4, respectively. } \label{figbeta}
\end{figure}

This paper is organized as follows: In the following section we
will present the functional integral formulation of the
Ginzburg--Landau free energy and the self-consistent gap equations
from a simple two band model.\cite{Lee09jpcm} Then we introduce
the anisotropy coefficient $\beta$ from the 4th order terms of the
GL free energy which determines the phase separation or
coexistence between the SDW and SC. In section III, we present the
detailed calculations of the multicritical point, the coefficient
$\beta$, and SDW and SC order parameters. We first show the
multicritical point in the $T-\delta$ plane in Fig.\ \ref{figbeta}
from which one can read off the nature of phase transitions. Then
we present three typical cases: phase separated C-SDW and SC in
Fig.\ 2, discontinuous SDW and SC phases in Fig.\ 3, and
coexisting IC-SDW and SC phases in Fig.\ 4. These results will be
commented in comparison with the experimental situations. Section
IV is for summary and concluding remarks.

\section{Formalism}

We take the following hamiltonian as with Ref.\
\cite{Vorontsov09prb} to describe the interplay between
superconductivity and antiferromagnetism.
 \ba
 \label{hamil}
H = \sum_{i,k,\sigma} \xi_{ik} c_{ik\sigma}^\dag
c_{ik\sigma} + H_{S}+H_{M}, \\
 H_{S} = \sum_{k,k'} V_{S} \left[ \ c_{1,k,\ua}^\dag
c_{1,-k,\da}^\dag c_{2,-k',\da} c_{2,k',\ua} + h.c. \right],
 \\
H_{M} = -V_{M} \sum_{k,\sigma} \sigma c_{1,k,\sigma}^\dag
c_{2,k+q,\sigma} \sum_{k',\sigma'} \sigma' c_{2,k'+q,\sigma'}^\dag
c_{1,k',\sigma'} ,
 \ea
where the $V_M$ and $V_S$ are the magnetic and superconducting
interaction strengths, respectively, and $q$ is the
incommensurability of SDW. $q=0$ and $\ne 0$ represent the C and
IC-SDW which is determined by maximizing the magnetic
susceptibility $\chi(q) $. See Eq.\ (\ref{ximq}) below.

The partition function of the Hamiltonian of Eq.\ (\ref{hamil}) in
the form of functional integral is given by\cite{Negele88}
($\tau$ is an imaginary time, and $\beta=1/k_B T$)
 \ba
\lefteqn{Z =} \nonumber \\ & & \int D[c_1^\dag,c_2^\dag,c_1,c_2]\,
e^{-\int_0^\beta \, d \tau \left[\sum_{k} (c_1^\dag
\partial_\tau c_1 + c_2^\dag \partial_\tau c_2)+H(\tau) \right] } .
\nonumber \\
 \ea
The next step is to employ the Hubbard-Stratonovich
transformations to decouple the electron-electron interactions of
$H_S$ and $H_M$. It is straightforward to obtain\cite{Lee09jpcm}
 \ba
e^{-\int_0^\beta d \tau H_S(\tau)} = \int
D[\Delta_1^*,\Delta_2^*,\Delta_1,\Delta_2] e^{-S_S}, \nonumber \\
 e^{-\int_0^\beta d \tau H_M(\tau)} = \int D[m^*,m] e^{-S_M},
 \ea
where
 \ba
S_{S} &=& \int d \tau \sum_{k} \, \left[ -\frac{ \Delta_1^*(\tau)
\Delta_{2} (\tau)}{V_{S}} - \left\{ c^\dag_{1,k, \uparrow}(\tau)
c^\dag_{1,-k,\downarrow}(\tau) \Delta_{2}(\tau) \right. \right.
\nonumber \\
 &+& \left. \left. c^\dag_{2,k, \uparrow}(\tau)
c^\dag_{2,-k \downarrow}(\tau) \Delta_{1}(\tau) +h.c. \right\}
\right ], \\
 S_{M} &=& \int d \tau \sum_{k} \, \left[ \frac{ m^*(\tau)
m (\tau)}{V_M} \right. \nonumber \\
 &-& \left. \left\{ m(\tau) \sum_\sigma \sigma
c_{1,k,\sigma}^\dag (\tau) c_{2,k+q,\sigma}(\tau) +h.c. \right\}
\right].
 \ea
$\Delta_{1/2}(\tau)$ are nothing but the (fluctuating)
superconducting order parameters. $\Delta_1$ and $\Delta_2$ are
taken as the opposite sign in accord with the $s_\pm$
pairing.\cite{Mazin08prl,Kuroki08prl} It is more convenient to
introduce
 \ba
\Delta = \sqrt{-\Delta_1 \Delta_2}
 \ea
as with Ref.\ \cite{Lee09jpcm}. Integrate out the fermions to get
 \ba
\label{partition2} \lefteqn{ Z[\Delta,m] = \int D[\Delta,m]
\,e^{-F},} \nonumber\\  F &=& \int_0^\beta d \tau \Big[ \int d\xi
\,\frac{1}{V_{S}} \Delta^*(\tau) \Delta(\tau) \nonumber \\ && +
\frac{1}{V_{M}} m^*(\tau)m (\tau) - \ln \det(
\widehat{M}_S+\widehat{M}_M) \Big ] . \nonumber \\
 \ea

We then make the saddle point approximation which is determined
by the condition that the first order functional derivative of
the action $F$ with respect to $\Delta$ and $m$ vanish.
 \ba
\left. \frac{\delta F}{\delta \Delta}\right|_{\Delta_{0},m_0}=
 \left. \frac{\delta F}{\delta m} \right|_{\Delta_{0},m_0} =0,
 \ea
where $\Delta_{0},m_0$ denote the saddle point values. These
functional derivatives can be computed using the following matrix
identity
 \ba \delta \ln \det M = \delta \mathrm{Tr} \ln M =
\mathrm{Tr}(M^{-1} \delta M).
 \ea
They are, of course, the usual gap equations:\cite{Lee09jpcm}
 \ba \label{gapeqdel1}
\lefteqn{\Delta = \sum_{\sigma=\pm}\frac12
\int_{-\omega_c}^{\omega_c} d\xi \lambda_{S} \frac{\Delta}{2
E_\sigma} \tanh\left(\frac{E_{\sigma}}{2T}\right),}
 \\ \label{gapeqm1}
\lefteqn{m = }\nonumber \\& & \sum_{\sigma=\pm}\frac12
\int_{-\omega_c}^{\omega_c} d\xi \lambda_M \frac{m}{2 E_\sigma}
\left(1+\frac{\sigma \delta } {\sqrt{\xi^2+m^2}}\right)
\tanh\left(\frac{E_{\sigma}}{2T}\right),
 \nonumber \\\ea
where the subscript $0$ was dropped,
 \ba
\lambda_S=N_F V_S,~~ \lambda_M=N_F V_M,
 \ea
where $N_F$ is the density of states at the Fermi level, and the
energies are given by
 \ba
E_\pm = \sqrt{\left(\sqrt{m^2 +\xi^2} \pm \delta
\right)^2+\Delta^2}. \label{enpm}
 \ea
We took $\xi_{1k}=-\xi+\delta$ and $\xi_{2k}=\xi+\delta$ for the
hole and electron Fermi surfaces,
respectively.\cite{Vorontsov09prb} The free energy is given as
follows:
 \ba\label{free}
\lefteqn{F =} \nonumber \\& & \frac{1}{\lambda_{S}} \Delta^* \Delta
+\frac{1}{\lambda_{M}} m^* m - 2T \sum_{\nu=\pm} \sum_k \ln
\cosh\left(\frac12 \beta
E_{\nu,k}\right). \nonumber \\
 \ea
We take $\omega_c$ as the unit of energy ($\omega_c=1$).

The second order terms are
 \ba\label{free2}
F^{(2)} = \left(\lambda_M^{-1}-\chi_M \right) m^* m
+\left(\lambda_S^{-1}-\chi_S \right) \Delta^* \Delta,
 \ea
where the pairing and magnetic susceptibilities are given by
 \ba
\chi_S(T) = \frac12 \int_{-\omega_c}^{\omega_c} d\xi \frac{1}{\xi}
\tanh\left(\frac{\xi}{2T}\right), \label{xis}\\ \chi_M(\delta,T)=
\frac12\int_{-\omega_c}^{\omega_c} d\xi \frac{1}{\xi}
\tanh\left(\frac{\xi+\delta}{2T}\right). \label{xim}
 \ea
When the doping is increased the magnetic transition may be
incommensurate, that is, $\chi_M$ can be maximum for IC wave vector.
For IC-SDW ($q\ne 0$), the magnetic susceptibility is given by
 \ba
\chi_M (\delta,T,q)= \frac12\left<
\int_{-\omega_c}^{\omega_c}d\xi\frac{1}{\xi-q\cos\theta}
\tanh\left(\frac{\xi+\delta}{2T}\right) \right>_\theta,
 \label{ximq}
 \ea
where $\left< ~~~ \right>_\theta$ implies the angular average.
Other technical details for IC case are collected in the appendix.
The $\chi_M (\delta,T,q)$ needs to be maximized with respect to
the $q$ for given $\delta$ and $T$. The magnetic transition
temperature $T_M$ and superconducting transition temperature
$T_c$ are determined by vanishing second order coefficients.

For comparison we also considered the equal sign $s$-wave pairing
to check how the interplay differs from the sign reversed
$s$-wave pairing. It is simple to see that the formula remain the
same except the energies $E_\pm$ of Eq.\ (\ref{enpm}). For equal
sign pairing, the energies are given by
 \ba
E_\pm = \sqrt{m^2 +\xi^2 +\Delta^2+\delta^2 \pm 2
\sqrt{(m^2+\xi^2)\delta^2 +m^2\Delta^2 } }. \label{enequal}
 \ea
The expressions for the $\chi_S$, $\chi_M$, and the 4th order
coefficients remain the same, except the $\beta'$ of cross term of
Eqs.\ (\ref{free4}) and (\ref{betaprime}). It will alter the
anisotropy coefficient $\beta$ of Eqs.\ (\ref{beta}) and
(\ref{coexist}) and the SC and SDW phases always separate for the
equal sign pairing.

Fig.\ 1 shows the magnetic transition temperature $T_M$ as a
function of $\delta$ determined by
 \ba
\chi_M (\delta,T_M,q)=\frac{1}{\lambda_M}
 \label{eqtm}\ea
for $\lambda_M =0.28,~ 0.4,~ 0.52,~ 0.68,$ and 0.8. Also shown in
the figure are $T_M^*$ and $T^*$ which are the boundaries between
C and IC-SDW, and between phase separation and coexistence. $T^*$
can be determined from the anisotropy coefficient $\beta$. It
would have been extremely difficult to map out the complete C/IC
SDW phase separation/coexistence with the SC phase without the
systematic approach using the coefficient $\beta$. It was computed
for general $q\ne 0$ IC-SDW as follows.

The nature of the transitions, that is, whether the magnetic and
superconducting transitions are continuous or discontinuous, and
whether $\Delta$ and $m$ coexist or not is determined by the 4th
order terms. Expansion of the GL free energy of Eq.\ (\ref{free})
up to the 4th order yields
 \ba\label{free4}
F^{(4)} =\beta_M |m|^4 +\beta_\Delta |\Delta|^4 +\beta'
|m|^2|\Delta|^2.
 \ea
All other combination terms vanish. The expressions of the
coefficients are collected in the appendix. If all coefficients
$\beta_M$, $\beta_\Delta$, $\beta'$ are positive, then the
magnetic and superconducting transitions as $T$ is reduced are
continuous. Depending on the relations among the coefficients as
will be discussed below, $m$ and $\Delta$ may coexist or repel
each other as the doping is varied below $T_c$. The negative
$\beta_M$, on the other hand, means discontinuous magnetic
transition as $T$ is reduced. This occurs in the vicinity of the
C-IC transition in the present model when $\lambda_M$ is large, as
will be discussed below.

In order to determine the phase coexistence/separation we follow
the standard statistical mechanics procedure and focus around the
multicritical point where
$T_M(\delta)=T_c$.\cite{Choi89prb1,Choi89prb2} We write
 \ba
\chi_S(T)-\lambda_S^{-1}= \frac{T-T_c}{T_c}, \label{tc} \\
\chi_M(T)-\lambda_M^{-1}= A_M \frac{T-T_c}{T_c},
 \ea
where $0<A_M<1$ is given by
 \ba
A_M = T_c \ \frac{\partial\chi_M (\delta_c,T_c,q_c)}{\partial T} \nonumber \\
 =
\frac12 \left<
\int_{-\omega_c}^{\omega_c}d\xi\frac{1}{\xi-q\cos\theta}\frac{\xi+\delta}{2T_c}
{\rm sech}^2\left(\frac{\xi+\delta}{2T_c}\right) \right>_\theta.
 \label{am}
\nonumber \\
 \ea

We now introduce the field $\eta$ to make the 2nd order terms
isotropic in the order parameter space.
 \ba
\eta = \sqrt{A_M}\ m
 \ea
and write
 \ba
F^{(2)} = \frac{T-T_c}{T_c} \left( |\eta|^2 + |\Delta|^2\right).
 \ea
Collecting the 2nd and 4th order terms we have
 \ba
F= \frac{T-T_c}{T_c} \left( |\eta|^2 + |\Delta|^2\right) +a|\eta|^4
+b |\Delta|^4 +c |\eta|^2|\Delta|^2, \nonumber \\
 \ea
where the coefficients are given by
 \ba
a= \frac{\beta_M}{A_M^2},~~b=\beta_{\Delta},~~
c=\frac{\beta'}{A_M}. \label{abc}
 \ea
The isotropic 2nd order terms suggest writing
 \ba\label{isotro}
|\eta|= r \cos\theta,~~|\Delta|=r \sin\theta.
 \ea

The free energy can now be written as
 \ba
\lefteqn{F =}\nonumber \\ & & \frac{T-T_c}{T_c}\ r^2 +r^4 \left[a
\cos^4\theta +
b\sin^4\theta +c\cos^2\theta\sin^2\theta \right]. \nonumber \\
 \ea
For
 \ba
a+b-c>0,
 \ea
the free energy takes the minimum with respect to $\theta$ when
 \ba
\cos^2\theta = \beta,
 \ea
where $\beta$ is given by
 \ba
\beta = \frac{b-c/2}{a+b-c}. \label{beta}
 \ea
We then have
 \ba
m\ne 0,~\Delta=0,~~ {\rm for}&~& \beta \ge 1, \label{sepa1} \nonumber \\
m= 0,~\Delta\ne 0,~~ {\rm for}&~& \beta \le 0, \label{sepa2} \nonumber\\
m\ne 0,~\Delta\ne 0,~~ {\rm for}&~& 0< \beta <1. \label{coexist}
 \ea

Eq.\ (\ref{coexist}), of course, represents the coexisting
region. It may be rewritten in terms of $a,~b,~c$ as
 \ba\label{coexist2}
a>c/2 ~{~\rm and~}~ b>c/2.
 \ea
Otherwise, the two orders phase separate and the first order
phase transition shows up as $\delta$ changes below $T_c$. The
coefficients $a,~b$, and $c$ can be calculated from Eqs.\
(\ref{free4}) and (\ref{abc}). Notice that the coexistence
condition of Eq.\ (\ref{coexist2}) is more restrictive than the
condition
 \ba\label{conti1}
ab>c^2/4
 \ea
given by Vorontsov $et~al.$\cite{Vorontsov10prb} The value of
$\beta$ determines the ratio of the two condensates by Eq.\
(\ref{isotro}) to give
 \ba
\Delta^2 = \frac{T_c-T}{2T_c} \frac{a-c/2}{ab-c^2/4},
 \nonumber \\
m^2 = \frac{1}{A_M^2} \frac{T_c-T}{2T_c} \frac{b-c/2}{ab-c^2/4}.
 \ea

The GL expansion is only valid close to the multicritical point.
As $T$ is lowered further below $T_c$, the order parameters become
larger and we need next-order terms in the GL expansion to
describe the critical phenomena with the same accuracy. This is
very tedious and inefficient. We therefore obtain the order
parameters $m$ and $\Delta$ self-consistently with numerical
iterations as explained before. In the following section we will
present the results.

\section{Results}

\begin{figure}
\includegraphics[scale=0.6]{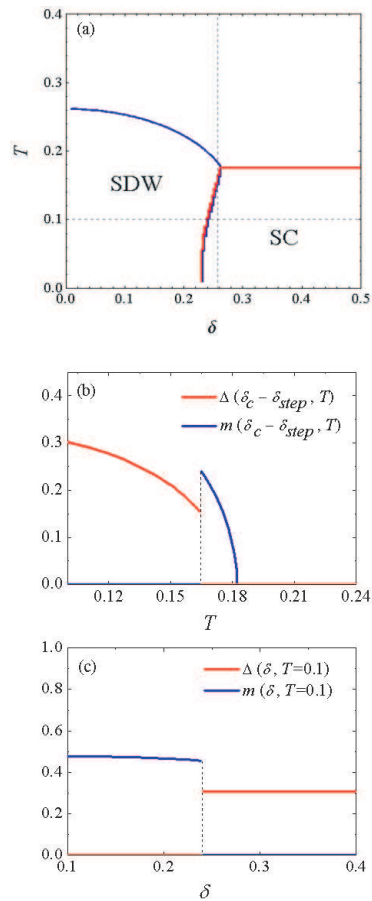}
\caption{(a) The phase diagram of the SDW and SC for
$\lambda_M=0.68$ and $\lambda_S=0.55$ where the tricritical point is
at the point A on the second line from the top in the Fig.\
\ref{figbeta}. It shows the phase separation between C-SDW and SC.
(b) The SC order parameter $\Delta$ and SDW order parameter $m$ as a
function of temperature at $\delta=\delta_c-\delta_{step}$ as
indicated by the vertical dashed line in (a). (c) The order
parameter $\Delta$ and $m$ as a function of $\delta$ at $T=0.1$ as
indicated by the horizontal dashed line in (a). } \label{figpoa}
\end{figure}

To map out the phase boundary between the SDW and SC phases we
first calculate $T_M(\delta)$ which is determined by $\lambda_M$
via Eq.\ (\ref{eqtm}). We considered the cases $\lambda_M =$ 0.28,
0.4, 0.52, 0.68, and 0.80 for representative values and
corresponding $T_M(\delta)$ are shown in Fig.\ \ref{figbeta}. The
incommensurability $q$ was computed by maximizing $\chi_M$ of Eq
(\ref{eqtm}) with respect to $q$ for given $T$ and $\delta$. $q=0$
and $\ne 0$ mean, respectively, the C-SDW and IC-SDW. The boundary
between $q=0$ and $q\ne0$ are marked on the $T_M(\delta)$ curves
by $T_M^*$. We found
 \ba\label{tmstar}
T_M^* = 0.56\ T_M^0
 \ea
in agreement with Ref.\ \cite{Vorontsov09prb}. This corresponds to
 \ba
\frac{ \delta}{2T_M^*} =0.958 \label{dotms}
 \ea
which can be calculated from $\chi_M$ of Eq.\ (\ref{ximq}) for
$q=0$.

We also computed the anisotropy coefficient $\beta$ along
$T_M(\delta)$ as the multicritical point is lowered along the
curve. For a given $\lambda_M$, a smaller $\lambda_S$ means that
the multicritical point where $T_M(\delta_c)=T_c$ is at larger
$\delta_c$ and smaller $T_c$. The computation of $\beta$ yields
that there exists the critical $T^*$ such that $m$ and $\Delta$
phase separate for $T_c>T^*$ and coexist for $T_c<T^*$. We found
 \ba\label{tstar}
T^*=0.35\ T_M^0.
 \ea
In Fig.\ \ref{figbeta}, the black color of each $T_M(\delta)$
curve represents the phase separated C-SDW and SC where
$T_M^0>T_c>T_M^*$, the cyan the phase separated IC-SDW and SC
where $T_M^*>T_c>T^*$, and the pink represents the coexisting
IC-SDW and SC phases where $T^*>T_c$. The coexisting SDW and SC
are possible only for IC cases for the sign changed pairing in
agreement with Ref.\ \cite{Vorontsov09prb}.

An interesting observation is made for large $\lambda_M$ such as
$\lambda=0.8$ of Fig.\ \ref{figbeta}. The coefficient $\beta_M$ of
Eq.\ (\ref{betaM}) of the 4th order SDW term becomes negative in
the vicinity of the boundary between the C and IC-SDW. It means
that the SDW transition as $T$ is reduced is discontinuous. See
the discontinuous changes of the order parameters as the
temperature is reduced as shown in Fig.\ 3(b). Incidentally, the
thermodynamic first order SDW transition was not reported in the
weak coupling calculation of Vorontsov $et~al$.\
\cite{Vorontsov09prb} To check the discontinuous C-SDW transition
as the mark B of Fig.\ \ref{figbeta} stands for, put $q=0$ in
$\beta_M$ given by Eq.\ (\ref{betaM}). Take the limit $\omega_c
\rightarrow \infty$ and put $x=\xi/2T$ and $y=\delta/2T$.
 \ba
 \beta_M = \frac{1}{(2T)^2} \int_{-\infty}^{\infty} \frac{dx }{x^3}
\left[ \tanh\left(x+y\right) - x\ {\sech^2}\left(x+y\right)
\right].
 \ea
The integral changes sign at $y=0.955$ which means that for
$\delta/2T_M >0.955$ the first order SDW transition occurs.
Comparing this with the $\delta/2T_M^*=0.958$ of Eq.\
(\ref{dotms}) means that the first order SDW transition intervenes
near the C-SDW and IC-SDW boundary. The self-consistent numerical
calculations indeed confirm this result as presented in Fig.\ 3.
In our calculation it does not show up when $\lambda_M$ is small
but begins to emerge when $\lambda_M \gtrsim 0.5$ probably because
the discontinuity for small $\lambda_M$ is too small. As $\delta$
increases further, however, the nonzero $q$ increases the
$\chi_M$, and IC second order transition preempts the first order
transition as shown in Fig.\ 1.

Then we consider three cases where the points marked by A, B, and
C on Fig.\ 1 are the multicritical point and calculated the phase
diagram in the $T-\delta$ plane. The SDW and SC order parameters
$m$ and $\Delta$ as a function of $T$ and $\delta$ are calculated
by solving the gap equations via numerical iterations. The results
like the phase separation/coexistence are fully consistent with
the GL expansion. Let us first consider the point A. The
computation of $\beta$ at the A point yields $\beta>1$
corresponding to the phase separation between C-SDW and SC. The
phase transition line below $T_c$ was calculated with the
numerical iterations of the gap equations of Eqs.\
(\ref{gapeqdel1}) and (\ref{gapeqm1}). The results are shown in
the Fig.\ \ref{figpoa}. Fig.\ \ref{figpoa}(b) and (c) present,
respectively, the order parameters $\Delta$ and $m$ as functions
of $T$ at $\delta=\delta_0-\delta_{step}$ and as functions of
$\delta$ at $T=0.01$ as indicated by the dashed lines. $m$ shows
the second order phase transition as a function of $T$, and the
$\Delta$ and $m$ show the first order phase transition as
functions of $\delta$ below $T_c$. This indicates that the SDW and
SC orders phase separate as the doping is varied as was mentioned
in the Introduction for the 1111 compounds.
La1111,\cite{Luetkens09naturemat} Sm1111,\cite{Drew09naturemat}
and Ce1111 compounds\cite{Zhao08naturemat} all showed the phase
separation as the doping was changed.

Now, let us turn to the case where the multicritical point is at
the point B corresponding to $\lambda_M=0.80$ and
$\lambda_S=0.55$. As alluded previously, the 4th order term of the
GL expansion $\beta_M$ at the point B becomes negative in this
case and the SDW transition as $T$ is reduced is discontinuous.
The phase diagram in the $T-\delta$ plane obtained by the
self-consistent numerical calculations of the gap equation is
shown in Fig.\ 3(a). The temperature dependence of the order
parameters $m$ and $\Delta$ is shown in (b). The $m$ exhibits the
abrupt onset at $T=T_M$. The experiments were reported to be the
continuous transitions for $m$ as a function of
$T$.\cite{Luetkens09naturemat} The transitions, however, could be
equally well described as a weakly first order as shown in the
present calculations. On the other hand, the abrupt change of
$T_M$ and $T_c$ as a function of doping was not seen in the
present model calculation. The experiments reported that the SDW
(SC) showed up in the orthorhombic (tetragonal) phase which
indicates the potential importance of the spin-lattice coupling.
The discrepancy between the experiments and the present
calculation could be due to the neglect of the spin-lattice
coupling.

\begin{figure}
\includegraphics[scale=0.6]{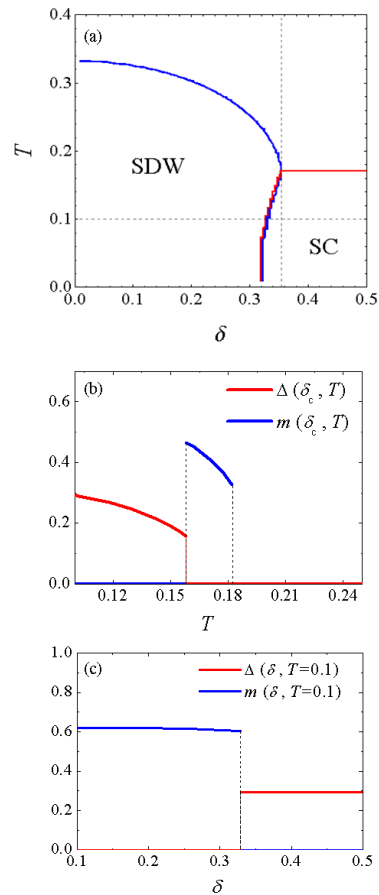}
\caption{(a) The phase diagram of the SDW and SC for
$\lambda_M=0.8$ and $\lambda_S=0.55$ corresponding to the point B
in the Fig.\ \ref{figbeta}. It exhibits the discontinuous change
of transition temperatures between C-SDW and SC at the critical
$\delta_c$. (b) The $\Delta$ and $m$ as a function of temperature
at $\delta=\delta_c$ as indicated by the vertical dotted line in
(a). (c) The order parameter $\Delta$ and $m$ as a function of
$\delta$ at $T=0.1$ as indicated by the horizontal dotted line in
(a). }
 \label{figpob}
\end{figure}

\begin{figure}
\includegraphics[scale=0.6]{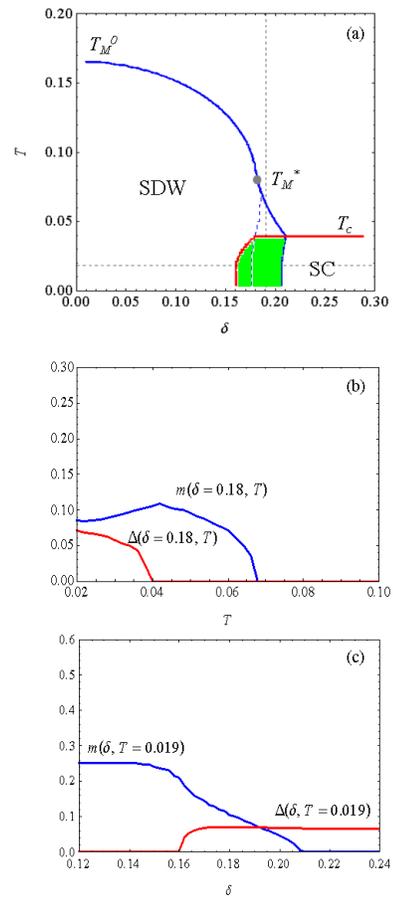}
 \caption{(a) The phase diagram obtained from the self-consistent
numerical iterations of the gap equations for
$\lambda_M=0.52,~\lambda_S=0.30.$ The tetracritical point
corresponds to the point C in the Fig.\ \ref{figbeta}. The green
shade in fig(a) represents the coexisting region of IC-SDW and
SC. The dashed line branching off from the $T_M^*$ represents the
boundary between C and IC-SDW. In figures (b) and (c), the order
parameters along the dotted lines of figure (a) are plotted. The
red line is $\Delta$ and the blue line is $m$ as a function of
temperature at $\delta=0.18$ and as a function of doping
$T=0.019$, respectively. All transitions below $T_c$ are
continuous. Note in figure (b) that $m$ decreases as $T$ is
lowered below the coexisting region because SC begins to gap out
the Fermi surface partially.} \label{figpoc}
\end{figure}

Fig.\ 4 shows the case where the tetracritical point is at the
point C corresponding to $\lambda_M=0.52$ and $\lambda_S=0.30$.
$0<\beta<1$ is this case which means the IC-SDW and SC phases
coexist microscopically, that is, both $m$ and $\Delta$ order
parameters are nonzero at the same point in the real space. The
phase diagram of Fig.\ 4(a) was obtained with the numerical
iterations of the gap equations of (\ref{energyq}),
(\ref{deltaq}), and (\ref{mq}). The order parameters change
continuously and the all the lines represent the second order
transition. The green shaded region of Fig.\ \ref{figpoc}
represents the parameter space where SDW and SC coexist
microscopically. Figure (b) plots $m$ and $\Delta$ as functions
of $T$ at $\delta=0.18$ indicated by the dotted vertical line in
figure (a). Note that $m$ decreases as $T$ is lowered below the
coexisting region because SC begins to partially gap out the
Fermi surface. Figure (c) shows $m$ and $\Delta$ as functions of
$\delta$ at $T=0.019$ indicated by the dotted horizontal line in
the plot (a). The SDW and SC order coexist in the region of $0.16
\lesssim \delta \lesssim 0.21$ in the low temperature limit. This
result nicely corresponds to the Co doped 122 compounds. As was
mentioned in the Introduction, Laplace $et~al.$\cite{Laplace09prb}
reported that the IC-SDW coexists with the SC on the atomic scale
in Ba(Fe$_{1-x}$Co$_x$)$_2$As$_2$. This result was later
confirmed by Julien $et~al$ on
Ba(Fe$_{1-x}$Co$_x$)$_2$As$_2$\cite{Julien09epl} and by Parker
$et~al.$ on NaFe$_{1-x}$M$_x$As (M=Co, Ni).\cite{Parker10prl}

\section{Conclusions}

In this paper, we examined the interplay between the spin density
wave and $\pi$ phase shifted superconductivity in the Fe pnictide
superconductors. We have obtained the phase diagram in the plane
of the temperature $T$ and chemical potential $\delta$ with the
combination of the Ginzburg-Landau expansion of the free energy
near the multicritical point and the self-consistent numerical
iterations of the gap equations. By calculating the multicritical
temperature $T_c$ as a function of the chemical potential
$\delta$ as shown in Fig.\ 1, we presented possible cases of phase
separation/coexistence among the commensurate SDW, incommensurate
SDW, and superconducting phases.

Then three typical cases were considered in more detail. The phase
separation of C-SDW and SC for $T_c>T_M^*$ was shown in Fig.\ 2.
The phase diagram in the $T-\delta$ plane, the $T$ and $\delta$
dependences of the SDW and SC order parameters were shown. In
Fig.\ 3, the discontinuous SDW--SC transition case was presented.
And in Fig.\ 4, the coexisting IC-SDW and SC for $T_c<T^*$ case
was presented.

In doing so, we employed a systematic way of determining the phase
separation or coexistence between SDW and SC orders from the 4th
order terms of the Ginzburg-Landau free energy expansion. We found
that if $T_c$ is larger than $T^*$ then the two orders phase
separate, but coexist for $T_c<T^*$ unless the first order
transition intervenes. Although the SDW transitions as
temperature is lowered are continuous for most of the parameter
space, they can become first order if the doping at the
multicritical point becomes large. Then the first order
transition intervenes in between the C-SDW and IC-SDW boundary.
This makes the phase boundaries more complicated than previously
reported as presented in Fig.\ 1.

Finally, we remark that the shapes of the electron Fermi surfaces
take quite different forms for different class of the pnictides.
The effect of the FS shapes on the phase coexistence was studied
in Ref.\ \cite{Vorontsov10prb}. Also, the effect of pairing
symmetry on the phase separation/coexistence will be interesting.
We are currently applying the present method to understand how
the phase coexistence/separation phenomenon depends on the
pairing symmetry.

\begin{acknowledgments}

This work was supported by by Korea Research Foundation (KRF)
through Grant No.\ NRF 2010-0010772.

\end{acknowledgments}

\appendix

\section{Incommensurate SDW}

In the appendix, we collect the technical details of the
Ginzburg-Landau expansion of the free energy for the
incommensurate SDW cases. The energy of Eq.\ (\ref{enpm}) above is
generalized for the non-zero incommensurability ($q\ne 0$) to
\begin{widetext}
 \ba\label{energyq}
E_\pm = \left\{ \left[ \sqrt{m^2+\left(\xi-q\cos\theta\right)^2}
\pm\left(\delta+q\cos\theta\right) \right]^2 +\Delta^2
\right\}^{1/2}, \label{eneq}
 \ea
and the SDW order parameter $m$ and superconducting order
parameter $\Delta$ to
 \ba\label{deltaq}
\Delta &=& \lambda_S \sum_{\sigma=\pm}\frac12
\left<\int_{-\omega_c}^{\omega_c} d\xi  \frac{\Delta}{2 E_\sigma}
\tanh\left(\frac{E_{\sigma}}{2T}\right)\right>_\theta,
 \\
m &=& \lambda_M \sum_{\sigma=\pm}\frac12 \left<
\int_{-\omega_c}^{\omega_c} d\xi \frac{m}{2 E_\sigma}
\left(1+\frac{\sigma \left(\delta+q\cos\theta\right) }
{\sqrt{\left(\xi-q\cos\theta\right)^2+m^2}}\right)
\tanh\left(\frac{E_{\sigma}}{2T}\right)\right>_\theta. \label{mq}
 \ea

The 4th order coefficients, $\beta_\Delta$, $\beta_M$, and
$\beta'$, of the Ginzburg--Landau expansion of the free energy
around the multicritical point of Eq.\ (\ref{free4}) may be
obtained by the derivative of the free energy of Eq.\
(\ref{free}) with respect to the order parameters. They are given
by
 \ba
\beta_\Delta=\frac{1}{4}\int_{-\omega_c}^{\omega_c}d\xi
\frac{1}{(\xi+\delta)^3} \left[ \tanh\left(\frac{\xi+\delta}{2T}
\right)- \frac{\xi+\delta}{2T}
{\sech^2}\left(\frac{\xi+\delta}{2T}\right) \right],
 \\
 \beta_M = \int_{-\omega_c}^{\omega_c} d\xi \left< \frac{1}{(\xi-q\cos\theta)^3}
\left[ \tanh\left(\frac{\xi+\delta}{2T}\right) -
\frac{\xi-q\cos\theta}{2T}
{\sech^2}\left(\frac{\xi+\delta}{2T}\right) \right]
\right>_\theta, \label{betaM}
 \\
 \beta'= \frac14 \int_{-\omega_c}^{\omega_c}
d\xi \left<\frac{1}{(\xi-q\cos\theta)(\xi+\delta)^2}\right>_\theta
\left[ \tanh\left(\frac{\xi+\delta}{2T}\right) -
\frac{\xi+\delta}{2T} {\sech^2}\left(\frac{\xi+\delta}{2T}\right)
\right]. \label{betaprime}
 \ea
\end{widetext}
The 4th order coefficients are then used to compute the
anisotropy coefficient $\beta$ using the equations (\ref{am}),
(\ref{abc}), and (\ref{beta}). The $\beta$ determines the phase
coexistence or separation by the condition of Eq.\
(\ref{coexist}).

\bibliographystyle{apsrev}
\bibliography{ref3}

\end{document}